# CRICTRS: EMBEDDINGS BASED STATISTICAL AND SEMI SUPERVISED CRICKET TEAM RECOMMENDATION SYSTEM

Prazwal Chhabra, Rizwan Ali and Vikram Pudi

International Institute of Information Technology, Hyderabad, India

## ABSTRACT

*Team Recommendation has always been a challenging aspect in team sports. Such systems aim to recommend a player combination best suited against the opposition players, resulting in an optimal outcome. In this paper, we propose a semi-supervised statistical approach to build a team recommendation system for cricket by modelling players into embeddings. To build these embeddings, we design a qualitative and quantitative rating system which considers the strength of opposition also for evaluating player's performance. The embeddings obtained, describes the strengths and weaknesses of the players based on past performances of the player. We also embark on a critical aspect of team composition, which includes the number of batsmen and bowlers in the team. The team composition changes over time, depending on different factors which are tough to predict, so we take this input from the user and use the player embeddings to decide the best possible team combination with the given team composition.*

## 1. INTRODUCTION

The advent of statistical modelling has contributed significantly to the success of teams and players in different sports. Different methods have been developed to evaluate player performances in different sports, but team sports pose a different challenge, as comparing two players of same nature and getting a suitable team against an opposition, is difficult. For example, in cricket [1], a team sport which is discrete in nature, comparing two players of same nature (comparing a batsman with another batsman, or a bowler with another bowler) in same and different teams is a complex task. Often, the players are compared based on their quantitative aspects like high scores, wickets taken and career averages (number of runs scored or conceded per dismissal) and teams are decided based on them only. The quantitative factors provide insights but miss some important aspects: Quality of Runs Scored: Two players who played against different oppositions (which are ranked differently) and performed similarly, will have similar statistics. In the mentioned case, the player who scored against better opposition, should be rated better. Quality of Dismissals: Dismissals of batsman with higher career average should be rated more than dismissals of batsman with lower career averages.

This paper tries to keep above two important aspects in mind and build a rating system called 'Quality Index of Player ($\phi$Player) which includes qualitative and quantitative aspects of player performance. Later, using $\phi$Player, we represent players as embeddings, to build a "Semi-Supervised Team Recommendation System". The embeddings obtained, describe the strengths and weaknesses of the players based on theirpast performances. While drafting a recommender system, factors like overall complexity and set of parameters to be considered, are a major factor and a system with high complexity won't be much useful for instantaneous results. If all the possible valid team combinations are considered from a pool of players, the complexity of that would still be polynomial, but very high. Proper selection of parameters along with considering orderings following some constraints can be useful for instantaneous results and can also be used for in-match results when match is not going as predicted. The method, although proposed for cricket, can also be extended to other sports with some modifications.

## 2. RELATED WORK

In literature, the player rating methods like A. Ramalingam [2], MG Jhawar et al. [3], S. Akhtar et al. [4], Margaret I. Johnston et al. [5] are quantitative in nature and give high weight to batsman with more batting averages (runs scored per wicket) and bowlers with lower bowling averages (runs conceded per wicket). Some studies include graphical representations to compare players (A. Kimber [6] proposed a graphical method to compares bowlers). Q. Zhou et al. [7] explains how team recommendations should work, considering the aspect of expanding teams and substituting team members. L. Li, H. Tong et al. [8] also explains team member replacement, considering skill and structure matching.

Also, the existing work on "Team Recommendation for Cricket" mainly rank players on statistical measures or some techniques like clustering, etc. Prakash, C. Deep [9] ranks players using a Clustering Algorithm based on different batting and bowling parameters. In S.B. Jayanth et al. [10], K-Means and SVM with RBF Kernel was used to recommend teams for 2011 World Cup. F. Ahmed et al. [11] maximizes the overall batting and bowling strength of a team by optimizing a Multi Objective Problem. NSGA-II algorithm was used to optimize the overall batting and bowling strength of a team and find team members in it.

## 3. PROPOSED SOLUTION

In this paper, we propose a qualitative and quantitative rating mechanism called 'Quality Index of Player $\phi$Player' which is used to build player embeddings. CRICTRS, a semi supervised team recommendation system, uses the player embeddings and recommends a team based on opponent's strengths and weaknesses. The system utilizes the weakness of the opponent and finds a similar player in our team to recommend against the opponent for each player in opposition. This process is done considering every player in the opposition. For representational purpose, a bipartite graph can be used, with opposition being on one side, and our players on the other.

### 3.1. Dataset

Cricsheet dataset [12] contains data of over 1400 international ODI matches, played between 2005 to 2019. For each match, ball by ball data is available, with following features: 'Inning', 'Over', 'Team Batting', 'Batsman', 'Non-Striker', 'Bowler', 'Runs-Scored', 'Extras', 'Wicket' and 'Dismissed Batsman'. Along with this, details like competing teams, date of match, venue of match, match and toss result are also available. Cricinfo [13] was used to validate the information across each match.

### 3.2. Player Rating System for Cricket

Improving upon existing methods we try to build a method that considers quality of runs and wickets while rating the players. A brief overview of CRICTRS is shown in Fig. 1.

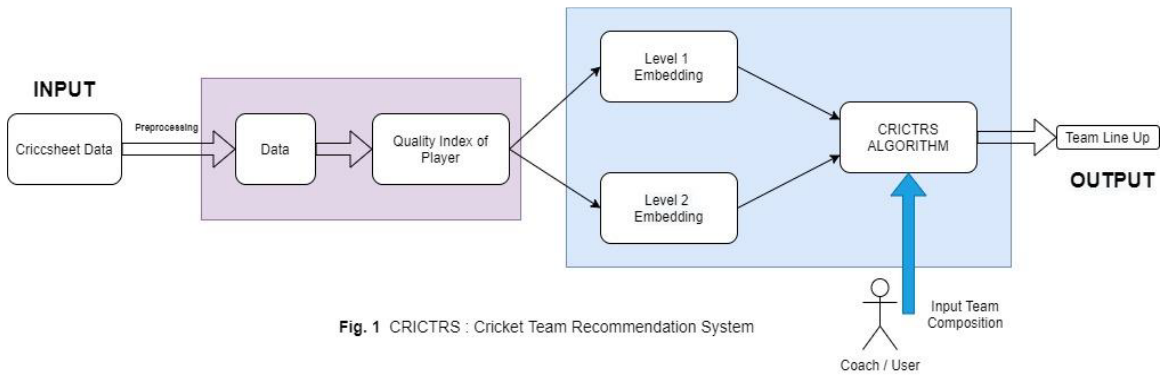

**Fig. 1** CRICTRS : Cricket Team Recommendation System

### 3.2.1. Modelling a Match

We take the idea from [2] and model each delivery as a Bernoulli trial. The two possible outcomes for each Bernoulli trial or a delivery are 'r' runs scored or a wicket, where 'r' is defined as average runs scored by a batsman per ball. To evaluate a batsman's individual performance in a team, we evaluate the performance of a team that contains 11 replicas of same player and calculate expected score by that team with 10 wickets in hand. Thus, a team with 11 replicas of batsman, on average, will score 300*r runs in a match, when the team does not lose any wicket in a 50 over (300 balls) ODI Match. But if a team loses 'w' wickets, where w<10, then the team will score (300-w) *r runs in that match. In case of an all-out, when team loses all the wickets, average runs scored will be (b-10) *r, where b is the number of balls the team faced in that match. Similarly, a bowler can be evaluated by replacing 'r' as average run conceded per ball and evaluating expected runs conceded by a team of 11 replicas of the bowler. Using above, the expected outcome of match can be written as: -

$$r = \frac{runs}{ball} \quad , \quad avg = \frac{runs}{wicket}$$

$$P(dismissal) = 1-p = \frac{1}{(balls\_per\_wicket)} = \frac{r}{avg}$$

$$E(runs) = E_{all\overline{out}}(runs) + E_{\overline{all}out}(runs)$$

$$E_{all-out}(runs) = \sum_{b=10}^{300} r*(b-10)*\left(\binom{b-1}{9}*p^{(b-1)-9}*(1-p)^9\right)*(1-p)$$

$$E_{\overline{all-out}}(runs) = \sum_{w=0}^{9} r*(300-w)*\binom{300}{w}*p^{300-w}(1-p)^w$$

$$E(runs^2) = \sum_{b=10}^{300} r*(b-10)*\left(\frac{b-1}{9}*p^{(b-1)-9}*(1-p)^9\right)*(1-p) + \sum_{w=0}^{9} r*(300-w)*\frac{300}{w}*p^{300-w}(1-p)^w$$

$$\sigma_{runs} = \sqrt{E(runs^2) - (E(runs))^2}$$

### 3.2.2. Quality of Runs and Dismissals

The approach in [2] is completely quantitative and misses an important aspect of quality of opposition. We replace the quantity metrics of 'r' and 'avg' used in [2] with our quality and quantity-based metric which is re-evaluated as follows: -

$$C_{batsman} = \text{Career Averarge of Batsman}$$

$$C_{bowler} = \text{Career Averarge of Bowler}$$

**Quality Metrics of Batsman**

$$\text{Quality of Dismissal } (\varphi \text{Dismisaal}) = \frac{C_{batsman}}{C_{bowler}}$$

$$\text{Quality of Run Scored}(\varphi \text{run}) = \frac{C_{batsman}}{C_{bowler}}$$

**Quality Metrics of Batsman**

$$\text{Quality of Dismissal } (\varphi \text{Dismisaal}) = \frac{C_{batsman}}{C_{bowler}}$$

$$\text{Total runs conceded} = \text{Runs Conceded} + \text{Extras}$$

$$\text{Quality of Run Scored}(\varphi \text{run}) = \frac{C_{bowler}}{C_{batsman}}$$

With this method we can consider some important aspects of the match, that are difficult to capture otherwise. These include: -

1) Dismissals of top order batsman matter more and as usually top order batsman have higher career average, thus a bowler who takes wickets of in form high average batsman is more rewarding than a bowler who takes wickets of tail-enders.

2) Extras were completely ignored by all previous metrics. Here if a bowler bowls more extras, then he might be under pressure, thus extras are also an important metric while considering bowlers. A bowler who gives away more extras provides greater risk to the team by giving away runs.

After re-evaluating the 'r' and 'avg' of players using above, we finally calculate the player rating represented by '**Quality Index of Player**' ($\varphi_{\text{Player}}$) which is evaluated as:

$$\text{Quality Index of Player } (\varphi_{\text{Player}}) = \frac{E(runs) - \varphi avg}{\sigma runs}.$$

On evaluating the results, we observe the following: -

1)      In [2] spinners and in general bowlers who bowled in the middle overs of the innings had higher rating than the bowlers who bowled in the powerplay and death overs, but our method regularises this as shown by the above examples. In above table, Harbhajan Singh and Yuvraj Singh had significantly higher ratings as compared to others by [2], but our method regularises the rating, and no such disparity is there.

2)      Also, there is a difference in rating by [2] and $\varphi$Player and some players are given higher rating by our method as compared to [2]. We believe that this is because these players performed better against strong oppositions, which should be valued more and [2] did not include this aspect of performance while rating the player.

3)　　　We compared performance of different players over the years and computed the rating at different stages of their career. Figure 2 shows rating of Virat Kohli's performance over years. The Quality Index and rating by [2] were normalised and plotted. Our method gave higher rating to his performance in 2012 as compared to 2016, which [2] rates as highest. On a closer look we see that the bowlers he faced in 2012 included experienced players like Lasith Malinga, Nuwan Kulasekara, Umar Gul, Saeed Ajmal, Clinton McKay, etc. While the bowlers he faced in 2016 included players like Jimmy Neesham, Josh Hazzlewood, Mitchell Santner, etc who were in the early stages of their career in 2016. Thus, we believe that the performance of Virat Kohli in 2012 should be rated more than 2016, as captured by our rating method also. A similar analysis is done for comparing Pat Cummins' performance over the years. We can see difference in ratings by [2] and our system, which is due to his performance against different oppositions.

Table 1. Rating of Batsmen

| Batsman | Rating By [2] | $\phi Player$ |
|---------|---------------|---------------|
| Virender Sehwag | 2.05 | 8.87 |
| Sachin Tendulkar | 4.88 | 11.45 |
| Gautam Gambhir | 3.47 | 12.95 |
| Yuvraj Singh | 2.66 | 7.08 |
| Mahendra Singh Dhoni | 6.80 | 9.50 |
| Yusuf Pathan | 1.23 | 3.07 |

Table 2. Rating of Bowlers

| Bowler | Rating By [2] | $\phi Player$ |
|--------|---------------|---------------|
| Zaheer Khan | 1.94 | 3.77 |
| Praveen Kumar | 1.75 | 3.40 |
| Ashish Nehra | 1.49 | 3.04 |
| Harbhajan Singh | 4.90 | 3.15 |
| Yusuf Pathan | 0.68 | 2.04 |
| Yuvraj Singh | 2.36 | 3.09 |

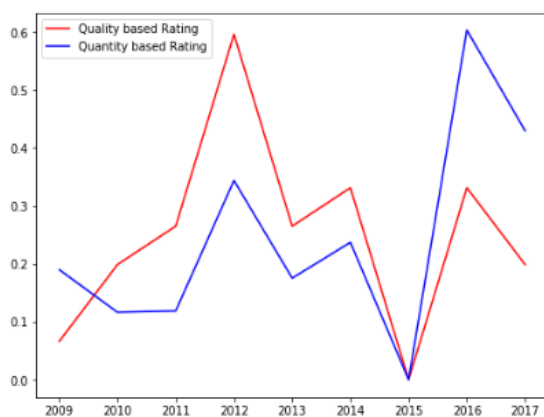

**Fig 2:** Performance comparison of Virat Kohli

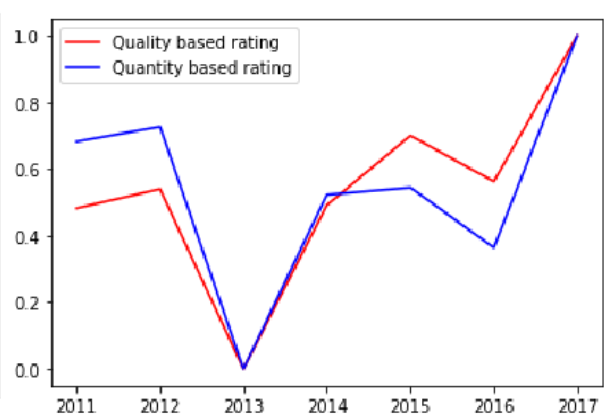

**Fig 3:** Performance comparison of Pat Cummins

## 3.3. Semi-Supervised Team Recommendation System

$\phi$Player proposed above helps us rate batsman and bowlers in cricket. But using rating directly to recommend a team can yield bad results as it does not capture the weaknesses and strengths of players, which are an important factor while forming a team against any opposition. Thus, we use an embedding based approach where we model each player as an embedding by comparing their performance against other players they face. The embeddings derived, capture the strengths and weaknesses of the players.

### 3.3.1. Player Representation: 2 Level Embeddings

We represent all the players as vectors. Every batsman is assigned a number $\in \{1, 2\ldots$ NBatsman$\}$, where NBatsman denotes the total number of batsmen. Similarly, every bowler is assigned a number $\in \{1, 2\ldots$ NBowler$\}$, where NBowler denotes the total number of bowlers.

Level 1 Embeddings: For a batsman, the Level 1 embedding is a vector, where index 'i' of the embedding gives $\phi$batsman against the bowler assigned the number i. Similarly, for a bowler, the Level 1 embedding is also a vector, where index 'i' of the embedding gives $\phi$bowler against the batsman assigned the number i. The

$\phi$Player at each index is calculated using batsman versus bowler data corresponding to that index, extracted from different matches in [12]. Also, all those indices in the embeddings are set to 0, for which versus data is not available, as those players did not face each other.

Level 2 Embeddings: The Level 2 embeddings are derived from Level 1 embeddings. For every batsman we compare $\phi$Player over his career against all the bowlers with the values at every index in the level 1 embedding representing batsman's performance against those bowlers. If the players have faced each other and there is a significant negative deviation in the performance of the batsman i.e. there is a drop in $\phi$Player for batsman, then the index is set to 0, otherwise it's set to 1. Similar process is followed for the bowlers and embeddings are constructed for bowlers also by comparing their performance against different batsmen with their overall performance. The level 2 embeddings, in a way, represent a batsman in terms of the bowlers he dominates, and the bowlers in terms of the batsmen they dominate.

In cricket, batsmen have weaknesses against bowlers. For example, a batsman may struggle against spin bowlers and play well against medium/fast bowlers. Thus, our embeddings capture the strengths and weaknesses of the players. Both levels of embeddings provide different views of player's strengths and weaknesses. Using the derived embeddings, we propose a 'Semi Supervised Recommender System' to suggest a team based on the given input and the opponent. We formulate our problem as: -

Suggesting a Team Given the player embeddings and team composition i.e. the number of players of each type (Batsman, Bowler and Wicket-Keeper), along with potential list of players in opposition, suggest a team of players best suited to face any combination of players of opposition.

### 3.3.2. Getting the Right Team Combination

Every team has a critical aspect of team composition, which includes the number of batsmen and bowlers in the team. The team composition changes over time depending upon the playing conditions, venue, opponent team, etc. No team combination is suited for all the conditions. The dataset [12] , only gives us ball by ball proceedings of the match, with no information regarding playing conditions. O. Alkan et al.

[14] builds a team based on the roles required for the opportunity and assigning the right person to fulfil it. In cricket, team composition decides the roles of different players. Thus, we propose a 'Semi Supervised' approach for our recommender system, where the coaches input desired team

composition. The team composition comprises of the number of batsmen, bowlers, and all-rounders in the team. Also, the coach inputs a set of players among which the team selection is to be made. Thus, our semi supervised recommender system combines the experience of a coach with our model to generate an optimum team combination.

### 3.3.3. Constraints on the Solution

The output is a set of (Player, Role), where Role is either batsman, bowler, all-rounder, or wicketkeeper. If it is being used for suggesting a team, then the following conditions should be satisfied: -

- Number of elements in the set >= 11
- Number of Players with the role wicketkeeper should be >= 1
- Number of Players with the role bowler and all-rounder should be >= 5

## 4. ALGORITHM FOR TEAM RECOMMENDATION

This section explains the algorithm used for CRICTRS in detail. Team configuration is taken as input for recommending players. Players are listed out using Level 2 embeddings against whom the opposing players are weak, from all countries. After that, for each opposition player, we check if there is a significant similarity in the Level 1 player embedding of a player from our team and any player in the above list of players against whom the opposition players have weaknesses. If so, then we can say that the player will outperform that opposition player. This approach is derived from collaborative filtering (explained in Y. Koren et al. [15]) as it uses an embedding based approach, although they were built without the use of supervised learning. Rather, we use a completely statistical approach to derive the embeddings.

We first adapt our solution to satisfy the constraints. Hence, the required number of wicketkeepers are selected first. Afterwards, the batsmen and bowlers are selected. Selection of these players depends on the opposition. For batsmen and wicketkeepers, the bowlers in opposition are considered, while bowlers are selected considering the batsmen in the opposition team.

### 4.1. Bipartite Graph Representation

A representation in the form of a bipartite graph can be constructed, where batsman is on one side, and bowlers are on the other, and vice versa. This representation is constructed based on the weakness of the opposition's players. A player of our team and the opposing team is connected by an edge, if the similarity between one of the players whom the player in the opposing team is weak, with our player, is below a threshold. Thus, we can construct such graph for batsmen and bowlers of our team, with bowlers and batsmen on the other side of the graph, respectively. Level 2 embeddings are used to construct the bipartite graph.

### 4.2. Ordering Players from Bipartite Graphs

Each edge on the bipartite graph has an edge weight equal to $\phi_{Player}$ against each other from Level 1 embeddings. For each player of our team, we compute $\delta$, which is defined as $\delta = \frac{Mean\ of\ Edge\ Weights}{Standard\ Deviation\ of\ Edge\ Weights}$ using the edges connected to the player node. The significance of selecting $\delta$ as the deciding parameter is that a player with high variance would have a lot of difference in the $\phi_{Player}$ against the player he/she dominates, thus providing greater risk to the team. Hence, stability across the players dominated is also considered in our algorithm. We sort the players in decreasing order of $\delta$ and make the selection.

### 4.3. Selection of Players from the Orderings

We construct the orderings of batsman, bowlers, and wicketkeepers. Wicketkeepers are selected first, as our recommendation should satisfy the constraints mentioned in section 3.3.3. While selecting the required number of batsmen, we check if they can bowl too and if so, whether they can bowl better than they can bat. So, we pick players with role batting all-rounder or bowling all-rounder. Similarly, bowlers and bowling all-rounders are picked.

## 5. EXPERIMENTS

CRICTRS involves a comprehensive embedding-based approach, where we represent a player as an embedding and select a team against the opposition using it. Different experiments were done to ensure the quality of individual components and a validation of overall results was also done.

### 5.1. Validating the Player Embeddings

The derived embeddings were validated by analysing the clusters of formed by them. Two of the obtained clusters from above are shown in table 3 and table 4.

Table 3. Cluster of similar Batsmen　　　　Table 4. Cluster of similar Bowlers

| Alastair Cook | | Saad Nasim |
|---|---|---|
| Marcus Trescothik | | Ajit Agarkar |
| Nathan Astle | | Khaled Mahmud |
| Hashim Amla | | Rubel Hossain |
| Virat Kohli | | Umesh Yadav |
| David Warner | | CR Braithwaite |
| | | Azhar Ali |

As we can observe from the above, the players in the same cluster, either have same playing style or they have dominated over similar kind of opposition. This validates our idea of modelling the players as embeddings, to capture the strengths, weaknesses, and other traits of the player, which is difficult to capture from statistics only.

### 5.2. Team Recommendation for Different Matches

We used CRICTRS to get teams for different matches. We obtained recommended team for South Africa, for the Australia v/s South Africa match on Oct 2, 2016. Using the Level 1 and Level 2 embeddings we obtained the bipartite graphs as shown in figure-4 and figure-5. The team recommended by CRICTRS, with a team composition of 5 batsman, 4 bowlers, 1 wicketkeeper and 1 bowling all-rounder is shown in table 5. On comparing with the team that played in the actual match, we see that CRICTRS suggests Hashim Amla in place of Behardien, while rest of the team is same. In the actual match, Behardien did not perform well. Also, when Hashim Amla played in the next match on Oct 5, 2016 against Australia then he performed significantly better than Behardien. Similarly, Indian team for the India v/s Pakistan match on 4th June 2017 was derived. As the two cricketing nations, do not play much cricket against each other, the versus statistical data is not available. Thus, CRICTRS provides an efficient team recommendation mechanism in such case by identifying similarities from the players opposition has already faced. The team recommended by CRICTRS, with a team composition of 3 batsman, 4 bowlers, 1 wicketkeeper, 1 batting all-rounder and 1 bowling all-rounder is shown in table 6.

Table 5. Recommended Players for Australia v/s South Africa

| Batsman | Bowler | Wicketkeeper | Batting All-Rounder | Bowling All-Rounder |
|---|---|---|---|---|
| Hashim Amla | Imran Tahir | Q De Kock | None | Wayne Parnell |
| Faf du Plesis | Kagiso Rabada | | | |
| David Miller | Dale Steyn | | | |
| JP Duminy | Andile Phelukwayo | | | |
| Rilee Rossouw | | | | |

Table 6. Recommended Players for India v/s Pakistan

| Batsman | Bowler | Wicketkeeper | Batting All-Rounder | Bowling All-Rounder |
|---|---|---|---|---|
| Rohit Sharma | Umesh Yadav | MS Dhoni | Kedhar Jadhav | Ravindra Jadeja |
| Virat Kohli | Jasprit Bumrah | | | Hardik Pandya |
| Shikhar Dhawan | R. Ashwin | | | |
| | B. Kumar | | | |

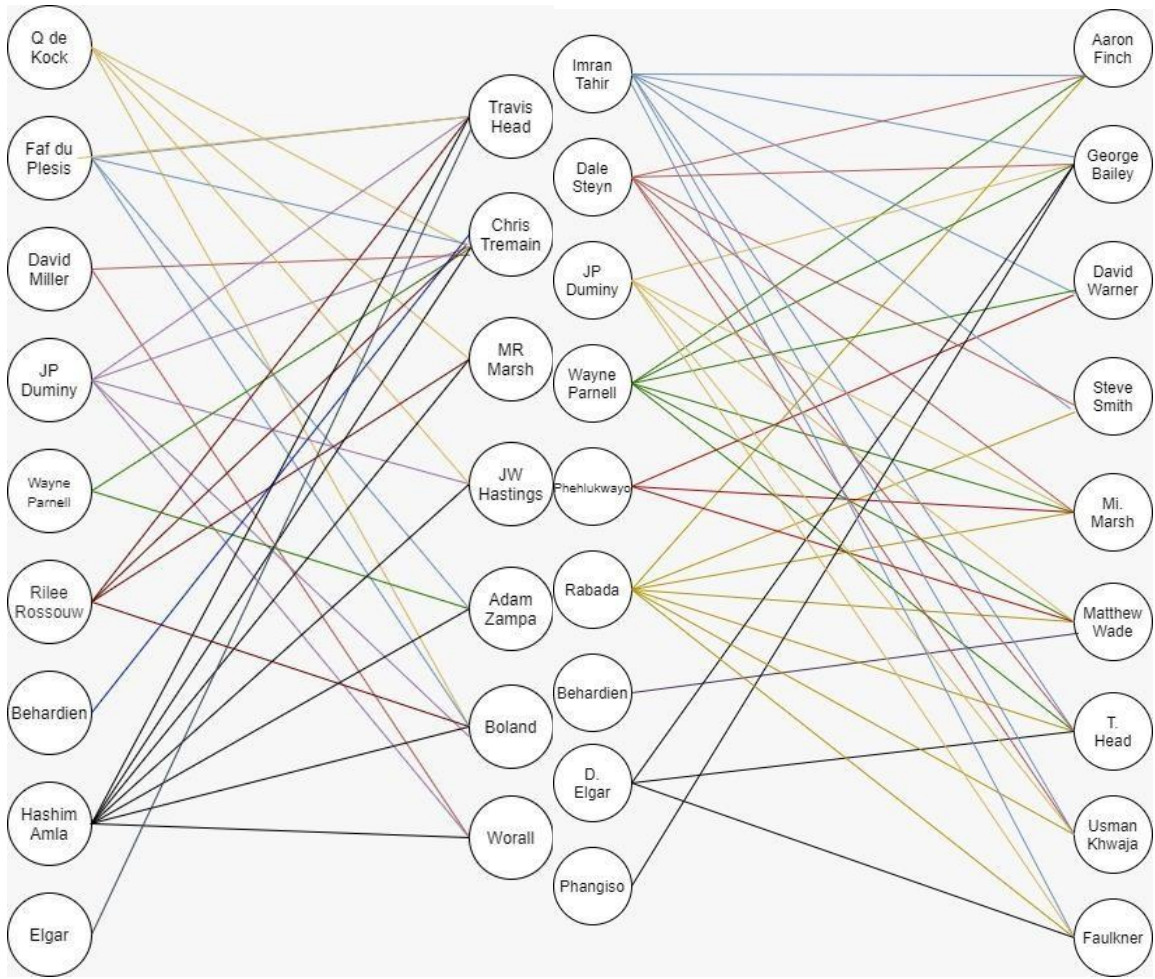

**Fig 4:** Bipartite Graph Representation of Batsman Bowler of South Africa against Australia

**Fig 5:** Bipartite Graph Representation of of South Africa against Australia

### 5.2.1. Team Line-Up similarity for ICC CWC - 2019

Comparisons were done between CRICTRS team recommendation results and actual team line-ups for ICC Cricket World Cup, 2019. The tournament had 48 matches in total, out of which 3 were abandoned due to rain, without a bowl bowled. For the other 45 matches, which had a winner, we compared the similarity in the line-ups generated by CRICTRS and the team line-ups in the actual match. The team composition for each team was kept similar to that in the actual match. The results obtained are shown in table 7: -

Table 7. Team Line-up similarity for ICC CWC-2019

|  | Team Line-up Similarity |
|---|---|
| Winning Team | 82.47% |
| Losing Team | 74.36 |

The low team line-up similarity with losing teams suggest that, CRICTRS recommends a different line-up for losing teams. Also, high similarity with winning teams validates that CRICTRS's recommended team line-up have high winning chances. Thus, above statistics prove useful in validating CRICTRS as a cricket team recommendation system.

## 6. CONCLUSION

CRICTRS provides a method to use historical data of different matches and recommend team based on opposition's strengths and weaknesses. CRICTRS models players as embeddings using the data and identifies weaknesses, strengths, and other traits of players. Cricket is an ever-evolving sport, with introduction of new rules, regulations, and technology. According to a new set of rules, the ICC allows players, who suffer concussions during a match, to be replaced in their team's playing XI [16]. However, the regulations emphasize on a 'like-for-like' replacement for the concussed player and that part remains to be under dark clouds [17]. Our embedding based approach can be used here, in finding and validating a 'like-for-like' replacement by checking the similarity in player embedding of the injured player and the replacement. In our analysis, we also tried to include domestic circuit players by using VORP ( Value Over Replacement Player )[18] theory from baseball and consider a run scored by a domestic level batsman against another bowler at domestic level, as 0.8 runs scored for the batsman and 1.2 runs conceded for the bowler. Thus, players across various levels can be compared with this and a uniform attribute is created.

Thus, CRICTRS can prove to be useful team recommendation tool and can help coaches and team management to decide their team line-ups against an opposition. Also, CRICTRS can also be extended to build a team recommendation system for other sports by modelling the sport game as a Bernoulli trial with appropriate outcomes. The $\phi$Player can be evaluated using the data, and the obtained rating system can be used to derive the embeddings in a similar manner.

## 7. ACKNOWLEDGEMENTS